\documentclass[runningheads]{llncs}
\usepackage{graphicx}
\usepackage{url}
\usepackage{graphicx}
\usepackage{amsfonts}
\usepackage[table,xcdraw]{xcolor}
\begin{document}
\title{Medical Diffusion: Denoising Diffusion Probabilistic Models for 3D Medical Image Generation}
%
%
\author{Firas Khader\inst{1} \and
Gustav Müller-Franzes\inst{1} \and
Soroosh Tayebi Arasteh\inst{1} \and
Tianyu Han\inst{2} \and
Christoph Haarburger\inst{3} \and
Maximilian Schulze-Hagen\inst{1} \and
Philipp Schad\inst{1} \and
Sandy Engelhardt\inst{4} \and
Bettina Baeßler\inst{5} \and
Sebastian Foersch\inst{6} \and
Johannes Stegmaier\inst{7} \and
Christiane Kuhl\inst{1} \and
Sven Nebelung\inst{1} \and
Jakob Nikolas Kather*\inst{8,9,10,11} \and
Daniel Truhn*\inst{1} 
}
\authorrunning{F. Khader et al.}
%
\institute{ 
Department of Diagnostic and Interventional Radiology, University Hospital Aachen, Germany \and
Physics of Molecular Imaging Systems, Experimental Molecular Imaging, RWTH Aachen University, Germany \and
Ocumeda AG, Germany \and
Artificial Intelligence in Cardiovascular Medicine, University Hospital Heidelberg \and
Department of Diagnostic and Interventional Radiology, University Hospital Würzburg, \and
Medical Clinic I, University of Mainz, Mainz, Germany \and
Institute of Imaging and Computer Vision, RWTH Aachen, Germany \and
Department of Medicine III, University Hospital Aachen, Germany \and
Else Kroener Fresenius Center for Digital Health, Medical Faculty Carl Gustav Carus, Technical University Dresden, Germany \and
Division of Pathology and Data Analytics, Leeds Institute of Medical Research at St James's, University of Leeds, UK. \and
Medical Oncology, National Center for Tumor Diseases (NCT), University Hospital Heidelberg, Germany
}
\maketitle              
\begin{abstract}
Recent advances in computer vision have shown promising results in image generation. Diffusion probabilistic models in particular have generated realistic images from textual input, as demonstrated by DALL-E 2, Imagen and Stable Diffusion. However, their use in medicine, where image data typically comprises three-dimensional volumes, has not been systematically evaluated. Synthetic images may play a crucial role in privacy preserving artificial intelligence and can also be used to augment small datasets. Here we show that diffusion probabilistic models can synthesize high quality medical imaging data, which we show for Magnetic Resonance Images (MRI) and Computed Tomography (CT) images. We provide quantitative measurements of their performance through a reader study with two medical experts who rated the quality of the synthesized images in three categories: Realistic image appearance, anatomical correctness and consistency between slices. Furthermore, we demonstrate that synthetic images can be used in a self-supervised pre-training and improve the performance of breast segmentation models when data is scarce (dice score 0.91 vs. 0.95 without vs. with synthetic data). 
The code is publicly available on GitHub: \url{https://github.com/FirasGit/medicaldiffusion}.
\end{abstract}
\section{Introduction}
The role of deep learning (DL) in medical imaging is steadily increasing \cite{aggarwal_diagnostic_2021}. A prototypical problem that can be solved by DL involves the classification of an image, i.e., the condensation of the high-dimensional data contained within the image down to a single class. The reverse action, i.e., the generation of medical images from low-dimensional non-image inputs is less often explored but has massive potential: synthetic images can be used to share protected data between sites or for educational purposes or can even be used to predict the progression of medical diseases in radiographs \cite{han_breaking_2020,han_image_2022}. These studies have been performed on two-dimensional (2D) images, but not three-dimensional (3D) volumes \cite{chen_synthetic_2021}. Yet, the most important diagnostic imaging modalities in modern medicine, magnetic resonance imaging (MRI) or computed tomography (CT), yield 3D data. Thus, the concentration on 2D data is a severe limitation which essentially ignores useful data which could be used to improve the performance and resilience of manual or automatic evaluation of these images. Hence, methods to generate synthetic 3D data are needed.\\

\noindent Previous works have employed generative adversarial networks \cite{kwon_generation_2019,eschweiler_3d_2021} (GANs), but this technique exhibits severe limitations: first, training of these models is difficult and mode collapse is a common problem \cite{thanh-tung_catastrophic_2020}, meaning that the neural network is unable to generate diverse samples. Second, the diversity of images generated by these models is limited even if no mode collapse occurs \cite{li_when_2021}. Third, GANs and similar models focus on the image domain only, and the generation of images from text or vice versa is not straightforward. Diffusion models on the other hand have had great success in the non-medical domain by being able to generate a wide diversity of images and by linking image and non-image data \cite{ramesh_hierarchical_2022,saharia_photorealistic_2022}. Despite their massively better performance, diffusion models have not been systematically used for 3D image generation in medicine.\\

\noindent In this work we examine if there is potential for diffusion models in medicine for the generation of 3D data. We present a new architecture for a diffusion model that works on the latent space and we train it on four publicly available datasets comprising data from a wide anatomical range: brain MRI, thoracic CT, breast MRI and knee MRI. We investigate if the images appear plausible to medical experts in a user study and quantitatively examine their diversity in terms of the structural similarity index. Finally, to bridge the gap to medical applications, we investigate whether pre-training on such generated synthetic images can contribute to improved segmentation models in limited-data settings. To foster future research and to establish a baseline to which other groups can compare, we make our code publicly available.

\subsection{Related Work}
Latent diffusion models have recently gained attention due to their success in generating high-quality images in the non-medical domain and by linking image data with text data \cite{rombach_high-resolution_2022}. There is continued interest in translating this success to the medical domain, since such models could be used for a wide variety of tasks, including data anonymization, education and training, data augmentation and discovery of new morphological associations \cite{kather_medical_2022}.\\

\noindent Medical image data however is partly - in the case of CT and MRI - more challenging than 2D image data due to its three dimensional nature. At the current point in time there is only one concurrent workshop presentation that we are aware of that has used diffusion models in the latent space to generate 3D MRI data based on a large database of brain scans \cite{pinaya_brain_2022}. Several other groups have worked on utilizing generative adversarial networks to generate 3D data \cite{han_synthesizing_2019}. In contrast, our approach can be seen as an extension of latent diffusion models, in which we append diffusion probabilistic models to the latent space of a VQ-GAN \cite{esser_taming_2021} to generate high-resolution 3D images. This approach has several benefits compared to the direct application of diffusion models to 3D data \cite{dorjsembe_three-dimensional_2022,kim_diffusion_2022}: (1) We can reduce the computational resources needed to train the model, since it is applied to a compressed latent space with smaller dimensionality. (2) While diffusion models excel at generating high-quality images, their sampling speed - especially in comparison to GANs - is fairly low. By sampling images in a lower dimensional latent space, we can decrease the time needed to generate new samples. (3) The latent space encapsulates more abstract information about the image content that covers whole areas of an image instead of pixel-level information, thereby making it accessible to advanced applications such as prediction of future image appearance irrespective of image rotation or translation \cite{han_image_2022}.

\section{Materials and Methods}
\subsection{Description of Dataset}
To demonstrate the performance and robustness of the Medical Diffusion model, we train it on four different publicly available datasets: The MRNet \cite{bien_deep-learning-assisted_2018-1} dataset contains 1,250 knee MRI exams from n=1,199 patients, each of which contains a scan of the axial, sagittal and coronal plane. For demonstrative purposes, we train our model exclusively on the sagittal T2 scans with fat saturation. The Alzheimer's Disease Neuroimaging Initiative (ADNI) \cite{petersen_alzheimers_2010} dataset contains brain MRI scans from n=2733 patients. The ADNI was launched in 2003 as a public-private partnership, led by Principal Investigator Michael W. Weiner, MD. The primary goal of ADNI has been to test whether serial magnetic resonance imaging (MRI), positron emission tomography (PET), other biological markers, and clinical and neuropsychological assessment can be combined to measure the progression of mild cognitive impairment (MCI) and early Alzheimer’s disease (AD). We train our model on 998 3D MP RAGE sequences labeled as cognitively normal (CN). Additionally, we evaluate our model on a breast cancer MRI dataset \cite{saha_machine_2018} acquired from 922 breast cancer patients, where we use the non-fat saturated T1-weighted sequences from each patient. To demonstrate the generalizability of our model, we also train the Medical Diffusion model to synthesize CT images. To this end, we used 1,010 (n=1010 patients) low-dose lung CTs from the Lung Image Database Consortium (LIDC) and Image Database Resource Initiative (IDRI) \cite{armato_lung_2011}. We also used an internal dataset of 200 (n=200 patients) T1-weighted breast MRIs with corresponding ground truth masks of the breast region to evaluate the use of synthetic breast images in a self-supervised pre-training approach.

\subsection{Data Preprocessing}
Knee MRI scans from the MRNet dataset were pre-processed by scaling the high-resolution image plane to 256x256 pixels and applying a histogram-based intensity normalization \cite{nyul_standardizing_1999} to the images. This procedure was performed by the dataset providers \cite{bien_deep-learning-assisted_2018-1}. Additionally, we center-cropped each of the images to a uniform shape of 256x256x32 (height, width, depth). The brain MRI sequences from the ADNI dataset were pre-processed such that the non-brain areas of the MRI images were removed. This process was done by the dataset providers. To allow comparisons between the diffusion model and GANs, we followed the approach by Kwon et al. \cite{kwon_generation_2019} and resized the brain MRIs to a shape of 64x64x64 before feeding them into the neural network. Images from the breast cancer dataset were preprocessed by first resampling all images to a common voxel spacing (0.66 mm, 0.66 mm, 3 mm) and then using the corresponding segmentation mask that outlined the breast to crop out a region of interest. The images were then split into two halves, such that the left and the right breast were on separate images. Finally, the images were resized to a uniform shape of 256x256x32. Accordingly, the lung CTs were first resampled to a common voxel spacing of 1 mm in all directions. Subsequently, the pixel values were converted to Hounsfield units and the images were center-cropped to a shape of 320x320x320 before being resized to 128x128x128. Images from all datasets were min-max normalized to the range between -1 and 1. Additionally, we augmented all datasets by vertically flipping the images during training with a probability of 50\%.

\subsection{Architecture}
The Medical Diffusion architecture is based on a two-step approach, where we first encode the images into a low-dimensional latent space and subsequently train a diffusion probabilistic model on the latent representation of the data. In the following, we first provide background information on vector quantized autoencoders, in particular the VQ-GAN \cite{esser_taming_2021}, and denoising diffusion probabilistic models \cite{ho_denoising_2020}:

\subsubsection{VQ-GAN}
To encode images into a meaningful latent representation, vector quantized autoencoders have shown to be a viable option as they mitigate the problem of blurry outputs in variational autoencoders \cite{van_den_oord_neural_2017,razavi_generating_2019}. They operate by mapping the latent feature vectors in the bottleneck of an autoencoder to a quantized representation, taken from a learned codebook. The VQ-GAN architecture proposed by Esser et al. \cite{esser_taming_2021} can be seen as a class of vector quantized autoencoders, whose image reconstruction quality is further improved by imposing a discriminator loss at its output. More precisely, images are fed into the encoder to construct the latent code $z_e \in \mathbb{R}^{(H/s) \times (W/s) \times (k)}$, where $H$ denotes the height, $W$ denotes the width, $C$ denotes the number of channels, $k$ denote the number of latent feature maps and $s$ denotes a compression factor. In the vector quantization step, the latent feature vectors are then quantized by replacing each one with its closest corresponding codebook vector $e_n$ contained in the learned codebook $Z$. The image is then reconstructed by feeding the quantized feature vectors into the decoder $G$. The learning objective is defined as a minimization of a reconstruction loss $L_{rec}$, a codebook loss $L_{codebook}$ and a commitment loss $L_{commit}$. As defined by the original authors, we use the perceptual loss as the reconstruction loss and use a straight-through estimator to overcome the non-differentiable quantization step. The commitment loss is defined as the mean squared error between the unquantized latent feature vectors and the corresponding codebook vectors. Note that the gradients are only computed for the continuous latent feature vectors to enforce a higher proximity to the quantized codebook vectors. The learnable codebook vectors are optimized by maintaining exponential moving averages over all the latent vectors that get mapped to it. In addition, a patch-based discriminator is used at the output for better reconstruction quality. To extend this architecture to support 3D inputs, we follow the approach by Ge et al. \cite{ge_long_2022} and replace the 2D convolutions by 3D convolutions. Additionally, we replace the discriminator in the original VQ-GAN model by a slice-wise discriminator that takes as input a random slice of the image volume and a 3D discriminator that uses the whole reconstruction volume as input. We also follow their approach in adding feature matching losses to stabilize the GAN training.

\subsubsection{Diffusion Models}
Diffusion models are a class of generative models that are defined through a Markov chain over latent variables $x_1 \cdots x_T$ \cite{ho_denoising_2020}. The main idea is that starting from an image $x_0$, we continuously perturb the image by adding Gaussian noise with increased variance for a number of $T$ timesteps. A neural network conditioned on the noised version of the image at timestep $t$ and the timestep itself is then trained to learn the noise distribution used to perturb the image so that the data distribution $p(x_{t-1} \mid x_t)$ at time $t-1$ can be inferred. When $T$ becomes sufficiently large, we can approximate $p(x_T)$ by the prior distribution $\mathcal{N}(\mathbf{0}, \mathbf{I})$, sample from this distribution and then traverse the Markov chain in reverse direction such that we can sample a new image from the learned distribution $p_{\theta}(x_0) := \int p_{\theta}(x_{0:T})dx_{1:T}$. The neural network used to model the noise is typically chosen to be a U-Net \cite{ronneberger_u-net_2015}. In order to support 3D data we replace the 2D convolutions in the U-Net with 3D convolutions. Additionally, we follow the approach by Ho et al. \cite{ho_video_2022} and only use convolutions on the high-resolution image plane (i.e. kernels of size $3 \times 3 \times 1$) followed by a spatial attention block on this high resolution plane (thus treating the depth dimension as an extension to the batch size) to increase the computational efficiency. The spatial attention block is then followed by a depth attention block, where the high-resolution image plane axes are treated as batch axes. 

\subsubsection{Putting It All Together}
In the first step, we train the VQ-GAN model on the whole dataset to learn a meaningful low-dimensional latent representation of the data. As the input fed into the diffusion model shall be normalized to the range between -1 and 1, we have to guarantee that our latent representation of the image is also in that range \cite{ho_denoising_2020}. Assuming that the vector quantization step in the VQ-GAN model enforces the learned codebook vectors to be close to the latent feature vectors before the quantization, we approximate the maximum of the unquantized feature representation by the maximum value in the learned codebook. Similarly, we approximate the minimum of the unquantized feature representation by the minimum value in the learned codebook. Thus, by performing a simple min-max normalization on the unquantized feature vectors, we achieve a latent representation with values close to the range -1 and 1. These can then be used to train a 3D diffusion model. We can then generate new images by running through the diffusion process in reverse direction, starting from noise sampled from a standard Gaussian. The output of this process is then quantized using the learned codebook of the VQ-GAN and subsequently fed into the decoder to generate the corresponding image. All models were trained on an NVIDIA Quadro RTX6000 with 24GB GPU RAM and took approximately 7 days for each model. More details about the training settings for each model can be found in Supplementary Table \ref{tab:hyperparams}.

\section{Results}
\subsection{Medical Diffusion Models can be Robustly Trained}
\begin{figure}
    \centering
    \includegraphics[width=\textwidth]{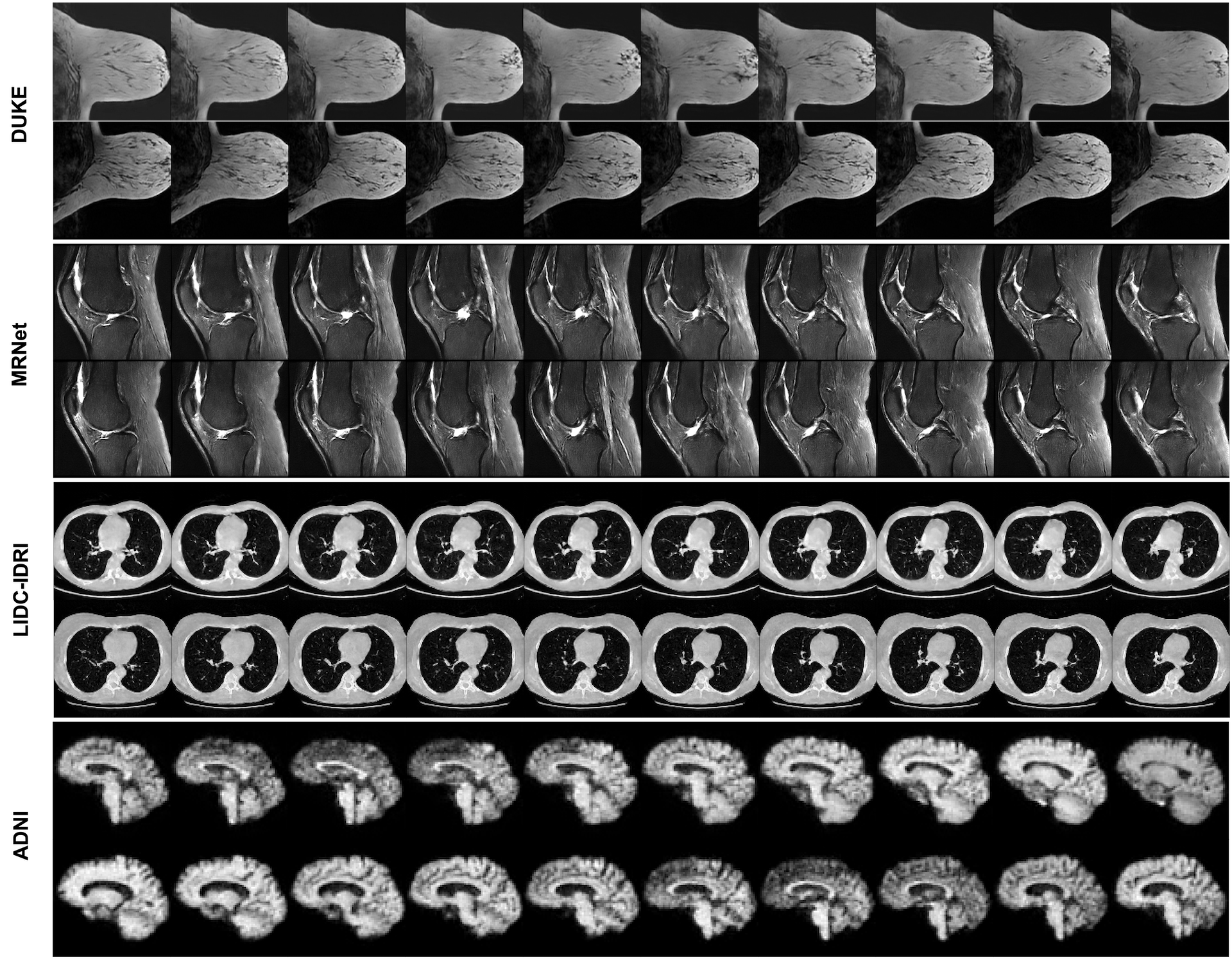}
    \caption{Visualization of synthetic data samples of the four public datasets used to train the medical diffusion model. Each row shows shows different neighboring z-slices of the same volume.To demonstrate the robustness to different resolutions, the model for the DUKE and MRNet dataset were trained with a resolution of 256x256x32 (height, width, depth), while the model for the LIDC-IDRI dataset is trained with a resolution of 128x128x128 and the model for the ADNI dataset is trained with a resolution of 64x64x64. }
    \label{fig:generated_samples}
\end{figure}

We trained the diffusion models on publicly available datasets from four different anatomical domains: brain MRI examinations from the Alzheimer's Disease Neuroimaging Initiative (ADNI), thoracic CT examinations from the cancer imaging archive (LIDC), breast MRI examinations from Duke University (DUKE), and knee MRI examinations from Stanford University (MRNet). To showcase the capability of our method even for small datasets, the four models was trained on 1,250 (knee MRI), 998 (brain MRI), 1,844 (breast MRI) and 1,010 (thoracic CT) images only. \\
Even though the datasets were comparatively small, each model converged and generated realistic synthetic images without fine-tuning any of the hyperparameters (Fig. \ref{fig:generated_samples}. In particular, we did not observe mode collapse in any of the training sessions. Additionally, the model architecture was capable of adapting to a wide range of resolutions, covering brain MRI with a resolution of 64x64x64 voxels, thoracic CT with a resolution of 128x128x128 as well as breast- and knee MRI with an anisotropic resolution of 256x256x32 voxels. In each of the four datasets, realistic three-dimensional data could be generated (Fig. \ref{fig:generated_samples}).

\subsection{Medical Diffusion Models can Generate High Quality Medical 3D Data}
\begin{figure}
    \centering
    \includegraphics[width=\textwidth]{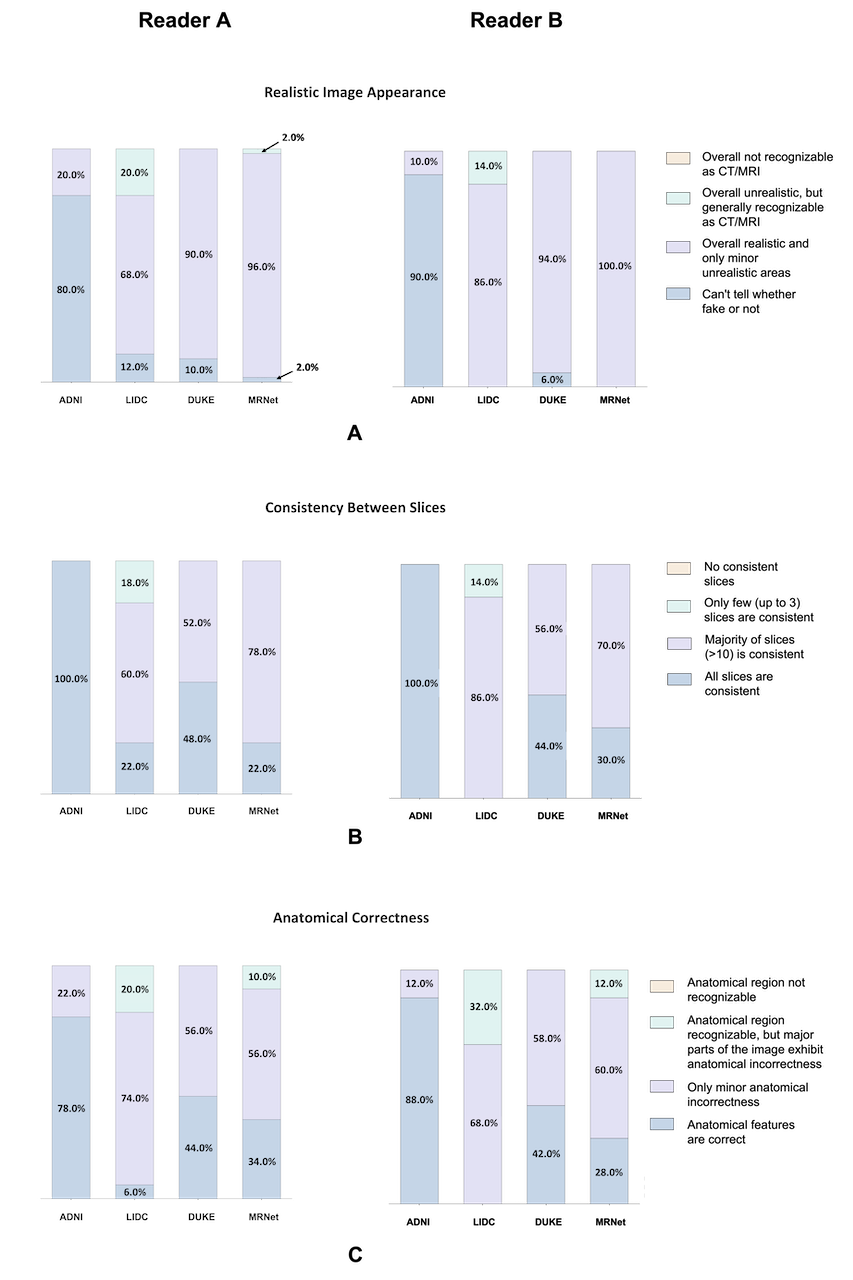}
    \caption{Quantitative evaluation of the image synthesis capabilities on the four datasets used to train the medical diffusion model. Two radiologists of 9 (Reader A) and 5 years (Reader B) of experience were tasked with evaluating a set of 50 synthetic images from each dataset on a Likert scale. The three categories used to assess the image are “Realistic Image Appearance”, “Consistency Between Slices” and “Anatomical Correctness”.}
    \label{fig:reader_study}
\end{figure}

We evaluated the synthetic images by a human expert along three different categories: 1) Quality of overall image appearance, 2) consistency between slices and 3) anatomical correctness. Two radiologists with 9 (Reader A) and 5 years (Reader B) of experience respectively were asked to rate 50 images from each of the four datasets on a Likert scale, see Table \ref{tab:categories}. The more experienced radiologist
rated 189 of 200 examinations to appear overall realistic with only minor unrealistic areas or better (50/50 for ADNI, 40/50 for LIDC, 50/50 for DUKE, 49/50 for MRNet). 191 of 200 examinations were rated to exhibit a consistency between slices in the majority of slices (50/50 for ADNI, 41/50 for LIDC, 50/50 for DUKE, 50/50 for MRNet) and 185/200 demonstrated only minor or no anatomic inconsistency (50/50 for ADNI, 40/50 for LIDC, 50/50 for DUKE, 45/50 for MRNet). Similar ratings were allocated by the radiologist with 5 years of experience (Fig. \ref{fig:reader_study}). Together these data show that our architecture can generate synthetic images which appear realistic to experts in the field.

\begin{table}[h]
\centering
\caption{Overview of the categories that were presented to the radiologist when tasked to assess the image quality. For each category, the radiologist was asked to select one out of the four options on a Likert scale. }
\resizebox{\textwidth}{!}{%
\begin{tabular}{l|l|l|l|l|}
\cline{2-5}
 & \textbf{Option A} & \textbf{Option B} & \textbf{Option C} & \textbf{Option D} \\ \hline
\multicolumn{1}{|l|}{\textbf{Realistic image appearance}} &
  \begin{tabular}[c]{@{}l@{}}Overall not \\ recognizable \\ as CT/MRI\end{tabular} &
  \begin{tabular}[c]{@{}l@{}}Overall \\ unrealistic, but \\ generally \\ recognizable as \\ CT/MRI\end{tabular} &
  \begin{tabular}[c]{@{}l@{}}Overall realistic and \\ only minor unrealistic \\ areas\end{tabular} &
  \begin{tabular}[c]{@{}l@{}}Can’t tell whether \\ fake or not\end{tabular} \\ \hline
\multicolumn{1}{|l|}{\textbf{Consistency between slices}} &
  \begin{tabular}[c]{@{}l@{}}No consistent \\ slices\end{tabular} &
  \begin{tabular}[c]{@{}l@{}}Only few (up to 3) \\ slices are \\ consistent\end{tabular} &
  \begin{tabular}[c]{@{}l@{}}Majority of slices \\ (\textgreater{}10) is consistent\end{tabular} &
  \begin{tabular}[c]{@{}l@{}}All slices are \\ consistent\end{tabular} \\ \hline
\multicolumn{1}{|l|}{\textbf{Anatomical correctness}} &
  \begin{tabular}[c]{@{}l@{}}Anatomical \\ region not \\ recognizable\end{tabular} &
  \begin{tabular}[c]{@{}l@{}}Anatomical region \\ recognizable, but \\ major parts of the \\ images exhibit \\ anatomical \\ incorrectness\end{tabular} &
  \begin{tabular}[c]{@{}l@{}}Only minor \\ anatomical \\ incorrectness\end{tabular} &
  \begin{tabular}[c]{@{}l@{}}Anatomical features \\ are correct\end{tabular} \\ \hline
\end{tabular}%
}
\label{tab:categories}
\end{table}

\subsection{The Dimension of the Latent Space is Important for High Quality Image Generation}
\begin{figure}
    \centering
    \includegraphics[width=\textwidth]{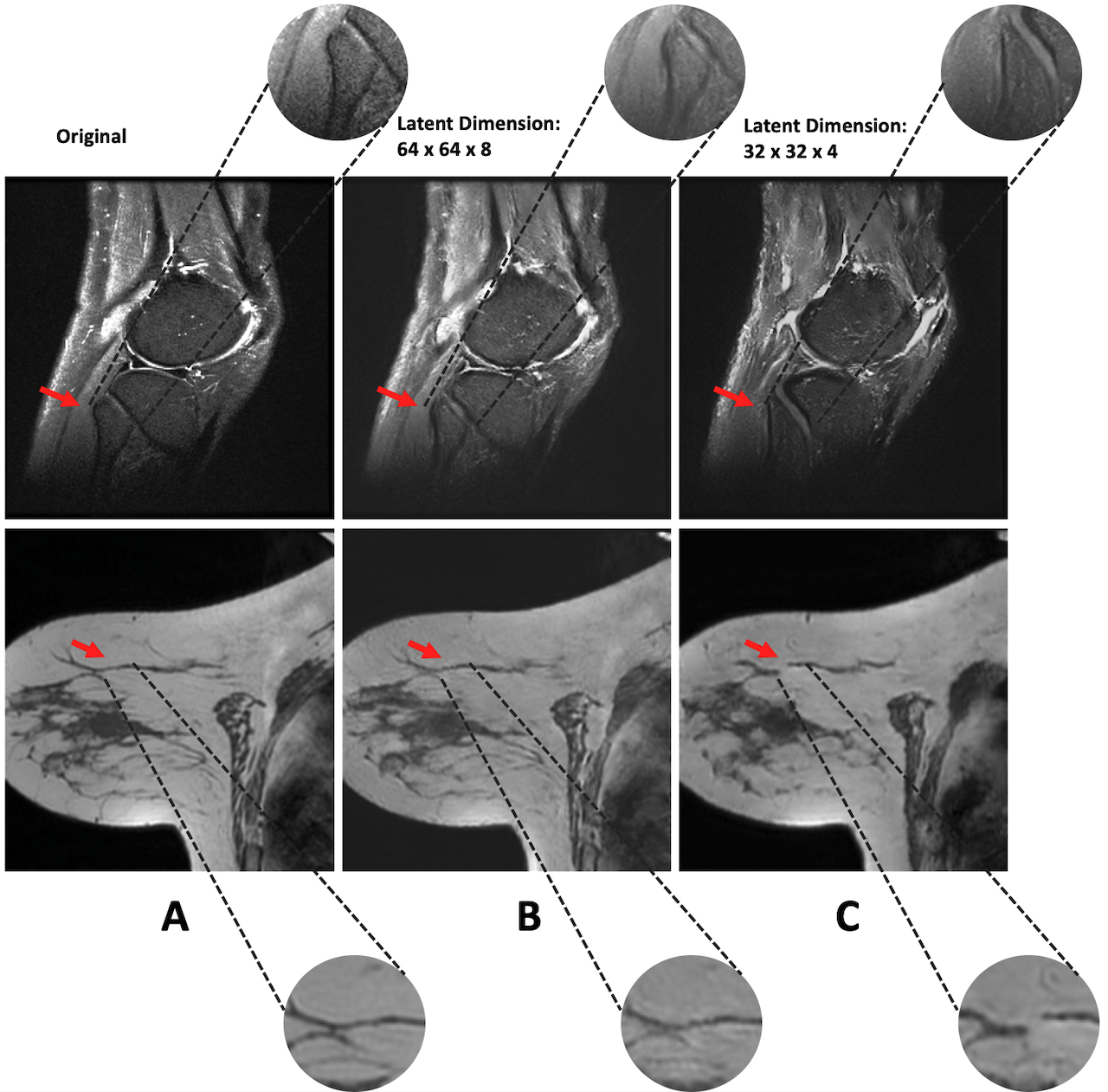}
    \caption{Comparison of the reconstruction quality of the VQ-GAN autoencoder when using different compression factors for two different samples. A latent dimension of 64 x 64 x 8 (i.e., a compression factor of 4 in each dimension) allows for a detailed reconstruction of the original image that conserves anatomical consistency. A compression factor of 8 (i.e., a latent dimension of 32 x 32 x4) distorts the fibular bone in knee MRI and the fibroglandular tissue continuity in breast MRI.}
    \label{fig:latent_space_vqgan}
\end{figure}

To analyze the effect of the latent dimension on image generation quality, we trained the VQ-GAN autoencoder with two different compression factors. We found that when compressing each of the spatial dimensions by a factor of 8 (i.e., images of size 256x256x32 have a latent dimension of 32x32x4), relevant anatomical features are lost (Fig. \ref{fig:latent_space_vqgan}). When training the VQ-GAN autoencoder with a smaller compression factor of 4 (i.e., images of size 256x256x32 have a latent dimension of 64x64x8), the anatomical features were reconstructed more accurately. For all four datasets, we found that a maximum compression factor of 4 (i.e., each dimension in the latent space was smaller by a factor of four as compared to the original dimension of the image) contained correct anatomical details as assessed by the radiological experts in a test set of 20 sample images for each dataset.

\subsection{Medical Diffusion Models outperform GANs in Terms of Image Diversity}
To compare our diffusion model to established GANs, we followed the work by Kwon et al. \cite{kwon_generation_2019} and chose a Wasserstein GAN with gradient penalty (WGAN-GP) as a baseline. Because we found divergent behavior in training the WGAN-GP with higher-resolution images, we restricted our comparison to the generated brain MRI images of size 64x64x64. We compared both models in terms of the multi-scale structural similarity metric \cite{wang_multiscale_2003} (MS-SSIM) by averaging the result over 1000 synthetic sample pairs of the same dataset. Higher MS-SSIM scores therefore suggest that the synthetic images generated by each model are more similar to each other, while a lower MS-SSIM score indicates the opposite. We found that the GAN model is not capable of generating diverse images as indicated by its high MS-SSIM score of 0.9996, resulting in synthetic images that are often identical. In contrast, the diffusion model achieved an MS-SSIM score of 0.8557, which is closer to the MS-SSIM score of the original data (0.8095). Together, these data show that diffusion models are able to generate more diverse samples representative of the original data distribution and that these models might therefore be better suited for follow-up projects, e.g. for training of classification models or for education.

\subsection{Synthetic Data can be Used to Train Neural Networks}
\begin{figure}
    \centering
    \includegraphics[width=\textwidth]{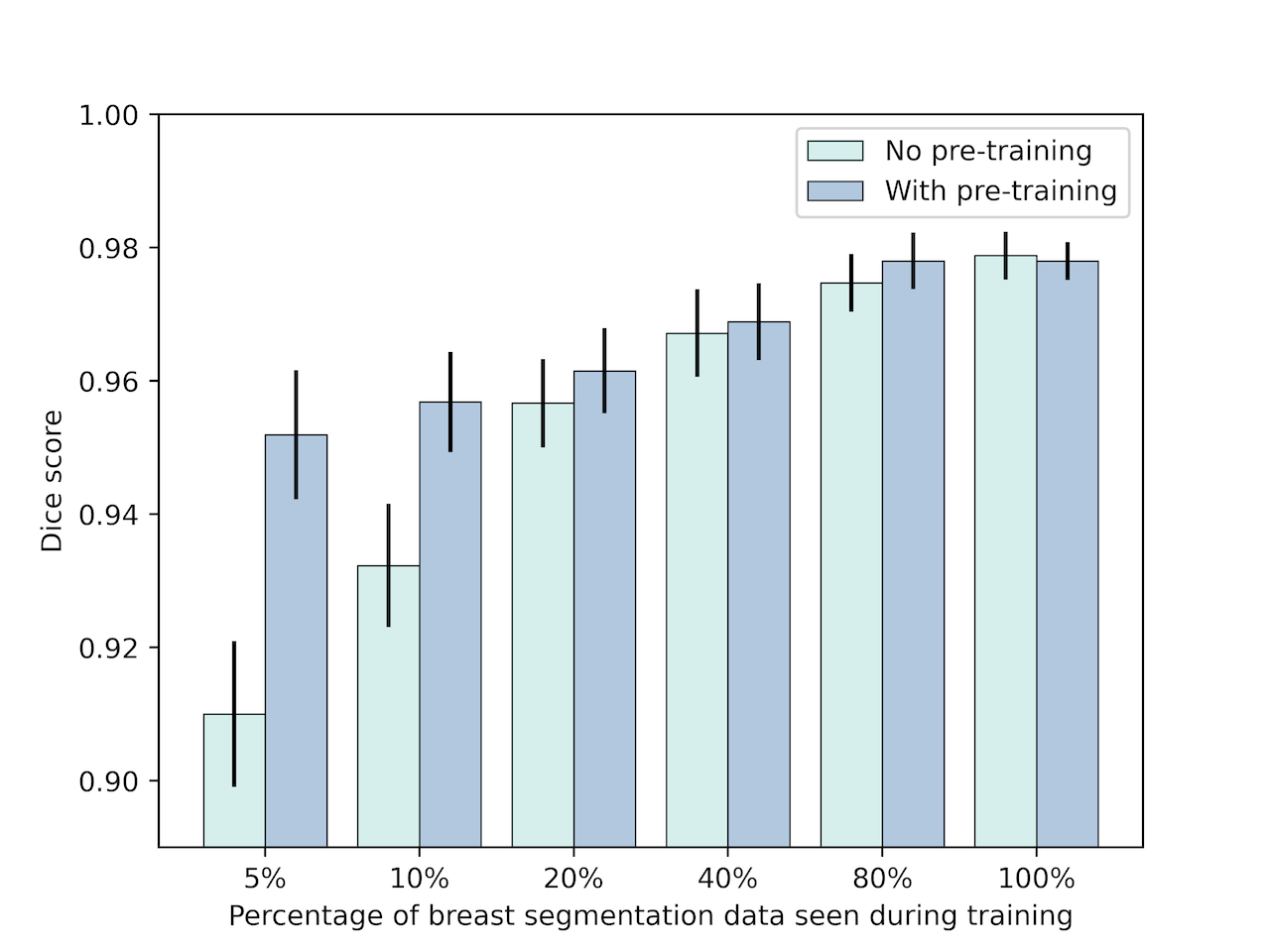}
    \caption{ Comparison of the performance of the Swin UNETR in terms of the Dice score for breast segmentation when supplied with synthetic image data (n=2000) during a self-supervised pre-training setting (“With pre-training”) and when no pre-training (“No pre-training”) is performed. After the optional self-supervised pre-training step, the model is finetuned with 5\%, 10\%, 20\%, 40\%, 80\% or 100\% of the available internal data with corresponding ground truth of the breast segmentation mask (n=200). We find that the synthetic data generated using a dataset of another institution can largely improve the Dice score on the internal dataset when only little data are available for training. }
    \label{fig:ssl_comparison}
\end{figure}

\begin{figure}
    \centering
    \includegraphics[width=\textwidth]{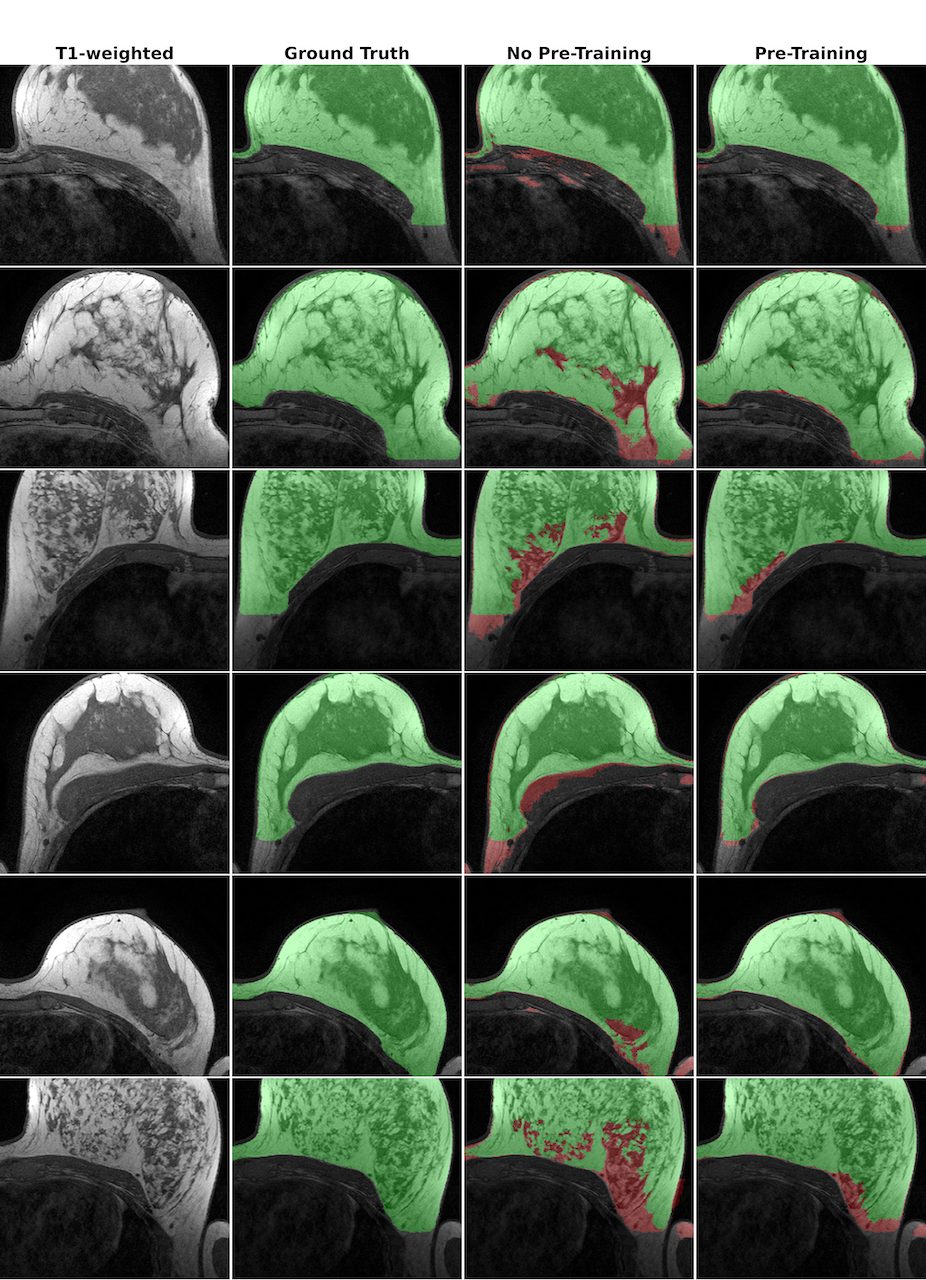}
    \caption{Visualization of the breast segmentation performance for three different cases (rows). The original image and the ground truth segmentation are shown in the first two columns. The third column shows the segmentation of the Swin UNETR neural network when 5\% of the available data from the internal dataset is used during training. The rightmost column shows the segmentation of the Swin UNETR when pre-trained in a self-supervised approach using 2,000 synthetic images generated using a dataset from another facility and then fine-tuned with 5\% of the available data from the internal dataset. Green areas denote a correct segmentation, whereas red areas denote a deviation from the ground truth.}
    \label{fig:segmentation_visualization}
\end{figure}

We evaluate the usability of synthetic data in a scenario where institution A wants to collaborate with institution B to increase the performance of a neural network without sharing any of the original data. To this end, we generated 2000 synthetic images using the diffusion model trained on the DUKE dataset and pre-trained a Swin UNETR \cite{tang_self-supervised_2022} in a self-supervised setting on the synthetic data. We then fine-tuned the pre-trained network with the available segmentation data from institution B to segment the breast region in the MRI scans. To showcase the performance increase in a limited-data setting, we performed multiple training runs in which we used an increasing amount of the available data from institution B (5\%, 10\%, 20\%, 40\%, 80\% and 100\%). For comparison, we trained the same neural network to perform the identical task, when no pre-training with synthetic data was performed. We found that pre-training with synthetic data from another institution can largely improve the segmentation performance in terms of the Dice score - especially in settings where available labeled training is small (0.91 without pre-training vs 0.95 with pre-training at 5\% available data, see Figs. \ref{fig:ssl_comparison} and \ref{fig:segmentation_visualization}).

\section{Discussion}
With the increase in quality of generative models in the non-medical domain, the synthesis of medical data becomes an attainable goal with potential applications in education, data anonymization, data augmentation and development of new DL algorithms \cite{han_image_2022,kather_medical_2022}. Diffusion models in particular have been shown to rival human abilities in image generation \cite{ramesh_hierarchical_2022,saharia_photorealistic_2022}. \\

\noindent In this work, we present the first large-scale evaluation of a latent diffusion model to both MRI and CT data. We demonstrate that such models can generate realistic 3D volumetric data that is both consistent in its synthesization of continuous 3D structures and that is capable of accurately reflecting human anatomy. We show that training of this complex data robustly converges for our models even if they are trained on comparatively small datasets of about 1,000 samples. \\

\noindent This is in contrast to GANs that usually require extensive hyperparameter tuning and large datasets for successful training. More importantly, even if GANs can successfully be trained, we find that our diffusion model is capable of more accurately encompassing the diversity of images encountered in medical practice. This is important for the use of such synthesized images in the development of AI methods. We additionally demonstrate a potential medical application of latent diffusion models by pre-training a segmentation model for human breast MRI examinations on synthesized data and show that such pre-training can help in rendering the segmentation models more robust. \\

\noindent Our work has limitations: first, we have evaluated our models on comparatively small datasets of about 1,000 examinations. This is partly by design - to showcase the possibilities of latent diffusion models when limited data is available - and partly due to limited computational resources. It can be expected that similar models can generate even more realistic images with higher resolution when trained on larger datasets \cite{pinaya_brain_2022}. Second, the generated 3D volumes do not have the full diagnostic resolution. This is due to the available public data, that is limited in resolution and does not in all cases reflect the state of the art in image resolution. We demonstrate that image quality and image resolution have a trade-off and that the compression factor for the latent space is crucial to arrive at realistic images. If large datasets become available for training of such diffusion models, e.g. through the use of federated approaches \cite{saldanha_swarm_2022}, our experiments indicate that image resolution can be scaled up, if the compression in the latent space is not chosen too high. \\

\noindent In summary, we have demonstrated that latent diffusion models are a superior method of generating synthetic 3D medical data in comparison to GANs and can form the basis for the development of AI methods on synthetic MRI or CT data.

\section{Data Availability}
We performed the diffusion model experiments on publicly accessible data to allow for other groups to reproduce and test our results. Only the breast segmentation model used to test the medical applicability of synthetic data relied on private data. This data is available upon request to the authors with a written cooperation and data protection agreement.\\
The LIDC-IDRI and the breast MRI (DUKE) datasets are available at the cancer imaging archive (TCIA) \cite{clark_cancer_2013}. The ADNI dataset is freely available at Image and Data Archive (IDA) \cite{crawford_image_2016}. The MRNet dataset is directly available from the dataset providers \cite{bien_deep-learning-assisted_2018-1}.

\section{Code Availability}
The code is publicly available on GitHub: \url{https://github.com/FirasGit/medicaldiffusion}

\section{Acknowledgements}
Data collection and sharing for this project was funded by the Alzheimer's Disease Neuroimaging Initiative (ADNI) (National Institutes of Health Grant U01 AG024904) and DOD ADNI (Department of Defense award number W81XWH-12-2-0012). ADNI is funded by the National Institute on Aging, the National Institute of Biomedical Imaging and Bioengineering, and through generous contributions from the following: AbbVie, Alzheimer’s Association; Alzheimer’s Drug Discovery Foundation; Araclon
Biotech; BioClinica, Inc.; Biogen; Bristol-Myers Squibb Company; CereSpir, Inc.; Cogstate; Eisai Inc.; Elan Pharmaceuticals, Inc.; Eli Lilly and Company; EuroImmun; F. Hoffmann-La Roche Ltd and its affiliated company Genentech, Inc.; Fujirebio; GE Healthcare; IXICO Ltd.; Janssen Alzheimer Immunotherapy Research \& Development, LLC.; Johnson \& Johnson Pharmaceutical Research \& Development LLC.; Lumosity; Lundbeck; Merck \& Co., Inc.; Meso Scale Diagnostics, LLC.; NeuroRx Research; Neurotrack Technologies; Novartis Pharmaceuticals Corporation; Pfizer Inc.; Piramal Imaging; Servier; Takeda Pharmaceutical Company; and Transition Therapeutics. The Canadian Institutes of Health Research is providing funds to support ADNI clinical sites in Canada. Private sector contributions are facilitated by the Foundation for the National Institutes of Health (www.fnih.org). The grantee organization is the Northern California Institute for Research and Education, and the study is coordinated by the Alzheimer’s Therapeutic Research Institute at the University of Southern California. ADNI data are disseminated by the Laboratory for Neuro Imaging at the University of Southern California.

\section{Author Contributions}
FK, DT and JNK conceptualized the study, performed and evaluated the experiments and wrote the first draft of the manuscript. MSH performed the reader study. TH, CH SE, BB, JS and SF advised in the development of the models. All authors corrected the first draft of the manuscript and agreed on the final manuscript.

\section{Competing Interests}
The Authors declare no competing financial or non-financial interests. For transparency, we provide the following information: JNK declares consulting services for Owkin, France; Panakeia, UK; and DoMore Diagnostics, Norway. Daniel Truhn declares consulting services for Nano4Imaging, Germany and Ocumeda, Switzerland.

\bibliographystyle{unsrt}
\bibliography{refs}

\newpage
\section{Supplementary Material}

\begin{table}
\renewcommand{\tablename}{Supplementary Table}
\setcounter{table}{0}
\centering
\caption{Hyperparameters used for training our Medical Diffusion Model.}
\resizebox{\textwidth}{!}{%
\begin{tabular}{|l|c|c|c|c|}
\hline
\rowcolor[HTML]{BDC1C6} 
 &
  \textbf{MRNet} &
  \textbf{ADNI} &
  \textbf{Breast MRI} &
  \textbf{LIDC-IDRI} \\ \hline
Modality       & MRI     & MRI     & MRI     & CT      \\ \hline
No. images     & 1,250   & 998     & 1844    & 1010    \\ \hline
\begin{tabular}[c]{@{}l@{}}Image Size \\ (height, width, depth)\end{tabular} &
  256x256x32 &
  64x64x64 &
  256x256x32 &
  128x128x128 \\ \hline
\textbf{VQ-GAN} &
  \multicolumn{1}{l|}{} &
  \multicolumn{1}{l|}{} &
  \multicolumn{1}{l|}{} &
  \multicolumn{1}{l|}{} \\ \hline
\begin{tabular}[c]{@{}l@{}}Compression rate \\ (height, width, depth)\end{tabular} &
  (4,4,4) &
  (2,2,2) &
  (4,4,4) &
  (4,4,4) \\ \hline
Codebook size           & 16384   & 16384   & 16384   & 16384   \\ \hline
Codebook dimensionality & 8       & 8       & 8       & 8       \\ \hline
Learning rate           & 3e-4    & 3e-4    & 3e-4    & 3e-4    \\ \hline
No. training iterations & 100,000 & 100,000 & 100,000 & 100,000 \\ \hline
Batch size              & 2       & 2       & 2       & 2       \\ \hline
\textbf{Diffusion Model} &
  \multicolumn{1}{l|}{} &
  \multicolumn{1}{l|}{} &
  \multicolumn{1}{l|}{} &
  \multicolumn{1}{l|}{} \\ \hline
No. training iterations & 150,000 & 150,000 & 150,000 & 150,000 \\ \hline
Learning rate           & 1e-4    & 1e-4    & 1e-4    & 1e-4    \\ \hline
Batch size              & 40      & 10      & 40      & 50      \\ \hline
Timesteps $T$           & 300     & 300     & 300     & 300     \\ \hline
\end{tabular}%
}
\label{tab:hyperparams}
\end{table}

\end{document}